# A cracking oxygen story: a new view of stress corrosion cracking in titanium alloys

Sudha Joseph[a,*], Paraskevas Kontis[b,¤*], Yanhong Chang[b,±], Yitong Shi[a], Dierk Raabe[a], Baptiste Gault[a,b,◊], David Dye[a,◊]

[a] Department of Materials, Royal School of Mines, Imperial College, Prince Consort Road, London SW7 2BP, United Kingdom.
[b] Max-Planck-Institut für Eisenforschung, Max-Planck-Str. 1, 40237 Düsseldorf, Germany.
* these authors contributed equally to the work.
± now at China Academy of Engineering Physics, Institute of Materials, Jiangyou, Sichuan, People's Republic of China
¤ now at Department of Materials Science and Engineering, NTNU Norwegian University of Science and Technology, 7034 Trondheim, Norway

## Abstract

Titanium alloys can suffer from halide-associated stress corrosion cracking at elevated temperatures e.g., in jet engines, where chlorides and Ti-oxide promote the cracking of water vapour in the gas stream, depositing embrittling species at the crack tip. Here we report, using isotopically-labelled experiments, that crack tips in an industrial Ti-6Al-2Sn-4Zr-6Mo alloy are strongly enriched (>5 at.%) in oxygen from the water vapour, far greater than the amounts (0.25 at.%) required to embrittle the material. Surprisingly, relatively little hydrogen (deuterium) is measured, despite careful preparation and analysis. Therefore, we suggest that a combined effect of O and H leads to cracking, with O playing a vital role, since it is well-known to cause embrittlement of the alloy. In contrast it appears that in $\alpha+\beta$ Ti alloys, it may be that H may drain away into the bulk owing to its high solubility in $\beta$-Ti, rather than being retained in the stress field of the crack tip. Therefore, whilst hydrides may form on the fracture surface, hydrogen ingress might not be the only plausible mechanism of embrittlement of the underlying matrix. This possibility challenges decades of understanding of stress-corrosion cracking as being related solely to the hydrogen enhanced localised plasticity (HELP) mechanism, which explains why H-doped Ti alloys are embrittled. This would change the perspective on stress corrosion embrittlement away from a focus purely on hydrogen to also consider the ingress of O originating from the water vapour, insights critical for designing corrosion resistant materials.

◊ Corresponding author: b.gault@mpie.de, d.dye@imperial.ac.uk

## 1 Introduction

Corrosion costs ~3% of global GDP[1], challenging lifetime extension and the sustainability of engineered structures[2]. Stress corrosion cracking is a major safety concern in many high-integrity structural materials, particularly corrosion and oxidation-resistant materials used in aggressive environments, such as Ni-base superalloys, stainless steels, titanium and zirconium alloys[3]. Usually such cracking is attributed to hydrogen ingress at stress concentrations sites, but hydrogen is difficult to detect at the nanoscale[4]. Hypothesised mechanisms involve grain boundary embrittlement[5] or hydrogen-induced localized softening and local ductility exhaustion[6].

Titanium alloys can be susceptible to chloride-associated cracking[6-8] in certain combinations of conditions, which is unfortunate as e.g. NaCl is ubiquitous in the environment, frequently being contained in the fine atmospheric dust ingested by jet engines and indeed, other industrial plant; in general, such corrosion-enhancing contaminants are difficult or impossible to completely avoid. In the commonly-held scheme of titanium stress corrosion cracking, the temperature and pressure used result in the evaporation of metal alloy chlorides $MCl_x$ (e.g. $TiCl_4$, $AlCl_3$, $SnCl_4$) and thermodynamically favours their reaction with water to produce HCl via e.g. $TiCl_{4(g)} + H_2O_{(g)} \rightleftharpoons TiO_{2(s)} + 4HCl_{(g)}$, rather than with oxygen which would produce $Cl_2$ gas via $TiCl_{4(g)} + O_{2(g)} \rightleftharpoons TiO_{2(s)} + 2Cl_{2(g)}$. The HCl can then react with metal at the crack tip to produce more alloy chlorides, e.g. $TiCl_3$, $AlCl_3$, $SnCl_4$, following a reaction such as $4HCl_{(g)} + Ti \rightleftharpoons TiCl_{4(g)} + 4[H]$. Rutile titanium oxide is one of the most efficient catalysts for the splitting of water[7], even more so when off-stoichiometry[8]. Thus, chloride and oxides can crack water from the vapour and deposit H into the crack and metal oxides on the crack mouth that then cause embrittlement of the alloy[9].

However, the influence of H on embrittlement has been inferred indirectly, often derived from post-mortem observations and/or by analogy to zirconium or to bulk H-charged samples, and is subject to intense debate in the community[10,11]. Kirchheim et al.[12] found that hydrogen lowered the energy barrier for the generation of dislocation loops. It is found experimental that a bulk content of 100ppmw H (~0.5 at%) is sufficient to embrittle Ti alloys[3]. Shih et al.[6] previously reported that the addition of H to a TEM chamber lowered the stress required for dislocation motion in mechanically loaded Ti foils, leading them to suggest that hydrogen enhances local plasticity (HELP) and that this is what leads to corrosion cracking[6]. This presupposes that H is in fact present from corrosion, which has not been observed directly. Parallels are often drawn with (single phase $\alpha$ or near-$\alpha$) zirconium alloys, in which the delayed hydride cracking phenomenon is a result of hydrogen migration to regions of high hydrostatic stress, which results in the precipitation of brittle hydrides that cause crack advance, even in static loading conditions at room temperature[25]. This is of concern, for example for the dry storage of spent nuclear fuel, where the hydrogen results from corrosion under elevated temperature operating conditions, where the solubility for H is elevated compared to room temperature.

In Ti alloys containing $\beta$-Ti, in contrast, H has ~50 at.% solubility in the continuous $\beta$-Ti matrix phase that acts as a sink, and H has a solubility of around 5 at.% in the $\alpha$ phase at elevated temperatures[3,13]. The hydrogen-induced decohesion mechanism is observed in some $\beta$-Ti alloys at high hydrogen concentrations and stress intensities [14,15]. On the other hand, if the stress effect on chemical potential were small, H may simply diffuse away into the bulk. Alloy design strategies have been developed to try and counteract the influence of H on stress-corrosion cracking[16]. However, observing and quantifying H within metals is extremely challenging. H lacks core electrons to excite, so conventional X-ray microanalysis cannot directly measure H in solution and therefore solute H embrittlement has yet to be directly observed.

Here, Ti-6Al-2Sn-4Zr-6Mo (wt.%) [Ti-10.8Al-0.8Sn-2.1Zr-3.0Mo (at.%)] used in jet engine compressor discs was supplied by Rolls-Royce plc. It was deformed under pressure and temperature conditions representative of the jet engine operating conditions. We used a combination of ion spectroscopies and electron spectroscopy and microscopy to examine the interactions of solutes with dislocations and compositional evolution during stress corrosion cracking. As a mass spectrometry technique, atom probe tomography (APT) has the intrinsic capacity to detect hydrogen and locate with sub-nanometre precision individual H atoms within a microstructure[17]. Isotope labelling is however necessary to avoid confusion with residual gases from within the microscope[4,18] and solute originally in the sample, so here we used $D_2^{18}O$. We have previously demonstrated that, borrowing from the biological sciences, cryogenic focused-ion beam specimen preparation is necessary to avoid spurious H ingress[19]. We show that beyond hydrogen, oxygen also penetrates inside the alloy, up to over 5 at%, leading to embrittlement. Only minor amounts of hydrogen/deuterium are detected, including specifically at crystalline defects. Our findings change the perspective on stress corrosion embrittlement away from a focus on hydrogen towards O originating from the water vapour, insights critical for designing corrosion resistant materials.

## 2 Materials & methods

### 2.1 Corrosion testing

Ti-6Al-2Sn-4Zr-6Mo (wt.%) [Ti-10.8Al-0.8Sn-2.1Zr-3.0Mo (at.%)] aero engine disc material received from Rolls Royce was investigated in this work, in a fully heat treated (slow cooled) and aged condition. Thin strips samples were cut with dimension 60 x 3.5 x 1.5mm. The samples were polished up to colloidal silica finish and a droplet of NaCl salt solution (0.1 g of NaCl in 100 ml deuterated water) was applied at the centre of each sample and allowed to dry before loading the sample in the bend rig.

An in-house designed 2-point bend rig was made of a titanium alloy, following the relevant ASTM standard[20]. The required stress was achieved by adjusting the holder span of the rig, S:

$$S = \frac{KtE(T)\sin\left(\frac{L\sigma}{KtE(T)}\right)}{\sigma} \text{------------- [1]}$$

where S is the distance between the two ends of the holder to achieve the desired stress σ, L and t are the length and thickness of the specimen, E(T) is the Young's modulus of the alloy at temperature T, and K is an empirical constant equal to 1.28. The true thickness of the sample was measured prior to loading, after polishing. Both the dimensions of the sample and S were measured using a calibrated digital vernier calliper with an accuracy of ± 0.1 mm. Tests were carried out under a surface stress of 650 MPa, below the yield point strength of the alloy (>950 MPa). The loaded sample was treated at a temperature of 450 °C for 100 hrs in a controlled environment wet rig. A continuous supply of $D_2^{18}O$ vapour (saturated at 80°C) with $^{16}O_2$ carrier gas at a pressure of 0.2 bar was maintained in the test rig. The test has been repeated on three samples under the same conditions.

### 2.2 Time of flight secondary ion mass spectroscopy

Time of flight secondary ion mass spectroscopy (ToF-SIMS) was conducted to investigate the diffusion of deuterium and oxygen near the crack. All the analyses were performed with a ToF-SIMS V instrument (IONTOF, Münster, Germany) equipped with a bismuth liquid metal ion gun (LMIG) with an incidence angle of 45°. The secondary ions were generated by the 25 keV $Bi^{3+}$ primary ion beam in the high current bunched mode (HCBM) for high mass resolution mass spectrometry. The ion beam current was 0.4 pA. SIMS spectra were captured from an area pre-ion polished near the crack to remove the oxide and contaminants from the surface.

### 2.3 Atom probe tomography

In order to avoid introducing hydrogen into the APT specimens and form artificial hydrides during their preparation, specimens were prepared under cryogenic conditions at -196°C. A bar was lifted out in a dual beam scanning-electron microscope / focused ion beam FEI Helios 600 using a protocol similar to the one outlined in ref. A GATAN C1001 stage was then mounted in the same microscope, and it was cooled to approx. -190C by a circulation of nitrogen gas cooled by liquid nitrogen. APT specimens were then sharpened, and left to warm back up to room temperature, transferred into the atom probe for analysis. The transfer time between the FIB and the atom probe stage maintained at 60K was minimised, and lasted less than approx. 4h. Specimens were analysed on a Cameca LEAP 5000 XR instrument operating in either voltage puling mode with 20% pulse fraction or laser pulsing mode with a repetition rate of 125khZ, pulse energy 30pJ and at a base temperature of 60K. Data reconstruction and processing was performed using the Cameca IVAS 3.8.4 software tool.

### 2.4 Electron microscopy

The specimens for TEM analysis were prepared using a focused ion beam (FIB) lift-out technique in a dual beam FEI Helios NanoLab 600 using a 30 kV Ga+ ion beam. TEM foils were lifted out from the crack tip on the thickness side of the sample and far away from the crack. To protect the area of interest, a gas injection system was used to deposit a platinum-containing protective layer. Samples were then made electron transparent by thinning down to a thickness of 150 nm. The samples were polished with 5keV and 2keV ions at the final stages to reduce the damage caused by the high energy Ga ion beam.

TEM and scanning TEM (STEM) analysis of the samples were carried out using a JEOL JEM-2100F TEM/STEM operated at 200 kV. The microscope was equipped with an Oxford Instruments X-Max 80mm$^2$ silicon drift detector for chemical analysis. EDS chemical analysis was carried out in STEM mode using a 1 nm probe. EELS was carried out in the same TEM equipped with a Gatan Tridiem imaging filter for EELS. The EELS energy resolution was between approximately 1 and 1.5 eV depending on the experimental conditions used, defined as the full width at half-maximum of the zero-loss peak (ZLP).

Finally, electron channelling imaging (ECCI) was performed after all the previous investigations. The oxide on the surface was removed by polishing and the surface was polished with a surface finish of 0.04µm colloidal silica. ECCI analysis of the deformed sample was performed on a Zeiss Merlin SEM equipped with a backscattered electron (BSE) detector, at an accelerating voltage of 20 kV and a probe current of 2 nA.

# 3   Results

## 3.1   Hot salt stress corrosion cracking

Figure 1a shows the 2-point bending set-up, at which specimens of Ti6246 were loaded at 650 MPa (~60% of the yield strength) with a drop of NaCl deuterated salt solution, in a stream of $D_2^{18}O$ vapour with $^{16}O_2$ carrier gas at a pressure of 0.2 bar and 450°C for 100 h. One of the resulting cracks, initiated where a salt crystal was located revealed by the presence of a crater, is shown in detail in **Error! Reference source not found.**b–e, illustrating the multi-scale nature of microstructural cracking. Note that the electron channelling contrast images in Figure 1c-e were captured in reversed contrast mode, which explains the bright contrast of the crack. A rutile $TiO_2$ film thinner than 100nm is observed at the surface of the specimen, indicated by red arrows in **Error! Reference source not found.**f. This film is far thicker than arises without the application of salt[21]. The microstructure, Figure 1c–e, consists of primary plates ~20 µm long of the hcp $\alpha_p$-Ti phase in a basketweave morphology, in a matrix of bcc $\beta$-Ti containing nm-scale secondary $\alpha_s$-Ti plates. The crack traverses this microstructure with only minimal deflection by the microstructural features, which is a typical so-called brittle behaviour usually reported for ductile metallic alloys under stress-corrosion cracking conditions[22].

## 3.2   Segregation of $^{18}O$ and Zr at dislocations.

We prepared a series of APT specimens every 2–3 µm ahead of the crack tip, with the needle-shaped specimen tip approx. 300–500nm below the exposed surface. The APT analysis, from ~5 µm ahead of the crack tip, shown in Figure 2**Error! Reference source not found.**a reveals almost no pick-up of D (<0.06 at.%). The composition of $^{18}O$ resulting from the water vapour in the gas stream is remarkably high at ~5 at.% of $^{18}O$, and as expected, oxygen segregates to the $\alpha$ phase, Figure 2b. The base alloy contained around 1800 ppmw $^{16}O$ (0.5 at.%), which is used as a solid solution strengthener of the majority $\alpha$ phase. Beyond around 2500 ppmw (0.8 at.%) O, titanium alloys are found to be brittle with negligible ductility, because O affects the dislocation behaviour. In the literature at least two possible mechanisms are discussed, (i) by enhancing $Ti_3Al$ precipitation result in a requirement for dislocations to travel in pairs, which supresses cross slip and results in localisation into slip bands, and (ii) as a consequence of strongly repulsive solute-dislocation interaction, meaning that gliding dislocations must clear the slip plane of interstitials, resulting in a pinning effect and in localisation[23]. Within $\beta$, Zr segregates along linear features, identified as dislocations in other alloy systems[24,25].  A compositional analysis from the core of this segregated region, Figure 2c, revealed almost 12 at.% $^{18}O$, but no D. There was also a slight increase in the H and $^{16}O$ signals, which can be attributed to the changes in the field evaporation behaviour at the linear defect.  Zr excess was also observed in the interface, Figure 2b **Error! Reference source not found.** & Figure 2d, believed to be associated with Zr segregation to interface (misfit) dislocations, which is also present in the undeformed sample[26]. No D or $^{18}$O excess was observed at the α/β interface.

.

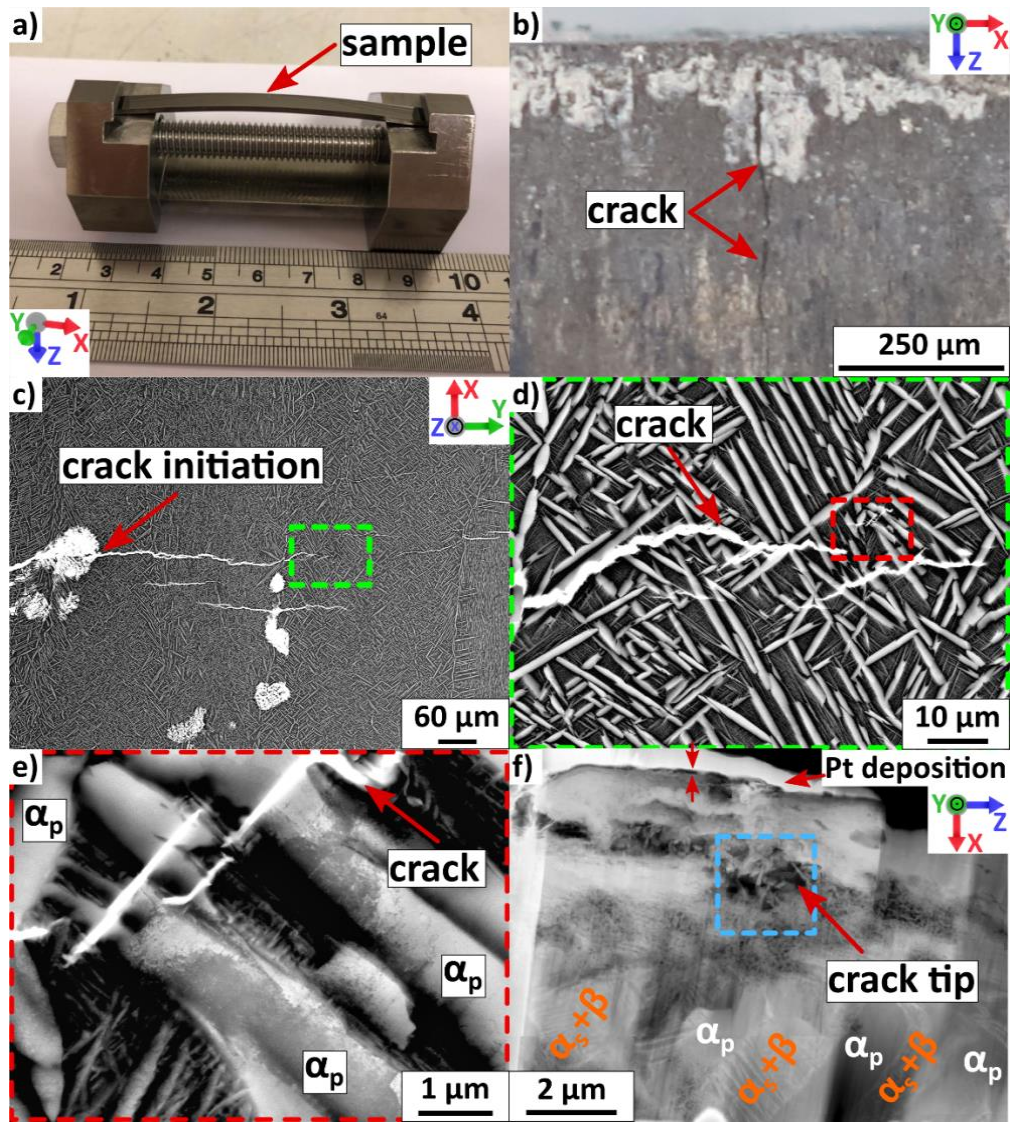

*Figure 1: Hot salt stress corrosion cracking in alloy Ti-6Al-2Sn-4Zr-6Mo. (a) Sample loaded in a two-point bending rig before heat treatment. (b) Optical image showing one of the cracks in the sample after testing. Inverse contrast ECCI micrographs showing: (c) a representative crack initiation site and crack growth within the specimen, (d)-(e) zoomed in detail of the crack tip from the crack initiated from where a salt crystal was located. (f) Zoomed-in detail of the crack tip area, from a FIB-lift out TEM foil imaged using HAADF-Scanning transmission electron microscopy. The blue dashed box in (f) denotes the region where the EELS analysis was performed in **Error! Reference source not found.**5.*

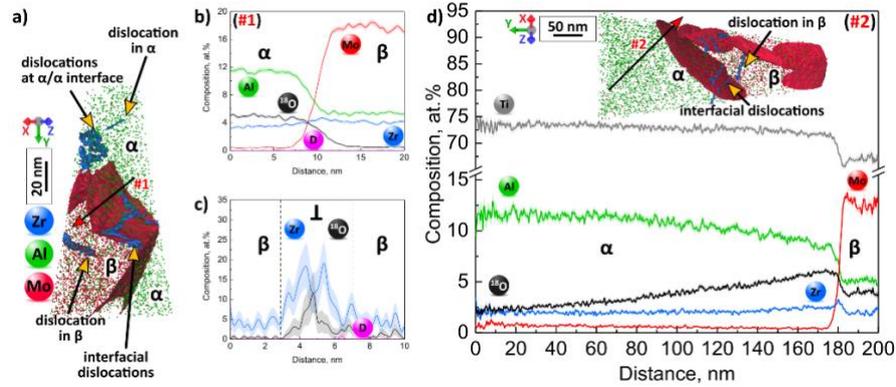

*Figure 2: Atom probe microanalyses. (a) APT reconstruction from a high dislocation density region ahead of a crack showing α/β and α/α interfaces and dislocations within the α and β phases and at the α/β interface. The dislocations are shown with an iso-composition surface at 5.9 at.% Zr and the α/β interface is shown with an iso-composition surface at 12 at.% Mo. (b) 1D composition profile across the α/β interface as indicated by arrow #1. (c) 1D composition profile across a dislocation in the beta phase. (d) 1D composition profile across the α/β interface in a different APT reconstruction as indicated by arrow #2, inset, from a high dislocation density region ahead of a crack. Error bars are shown as lines filled with colour and correspond to the 2σ counting error.*

### 3.3 Plasticity-assisted compositional evolution.

The presence of $^{18}$O at dislocations in β is indicative of the possibility of pipe diffusion of oxygen[27]. The composition profile in Figure 2**Error! Reference source not found.Error! Reference source not found.**d was obtained from an APT analysis obtained almost at the crack tip. This evidences a strong $^{18}$O ingress into α, which extends over a very substantial distance – over 150 nm, from the α/β interface. The high concentration of $^{18}$O in β is a strong indication that the continuous network of β helps with the penetration of oxygen within the alloy, since O has a much higher diffusivity in β compared to α[28]. The presence of a high density of dislocation in this region can help increase the amount of O accommodated within β, especially in light of Figure 2**Error! Reference source not found.**c, and facilitate O transport via pipe diffusion[27,29] through solute dragged by moving dislocations[30,31]. These mechanisms may facilitate the ingress of O into α.

We also performed APT prior to stress-corrosion testing. The composition of the α phase is notably different after corrosion cracking at the crack tip, where the Ti content decreases from 83 to 81 at.%, Sn from 0.8 to 0.1 at.% and Mo from 0.45 to 0.2 at.% whilst the Zr content increases from 3.2 to 5.6 at.% (Figure 3). In previous work, we have observed the formation of alloy chlorides in the interaction layer with the salt particle consistent with this result, and long-range transport of solutes in highly plastically deformed regions has also been observed in Ni superalloys[32,33]. Increases in Zr content in TiZr binary alloys have previously been associated with an increase in hardness and drop in toughness[34,35].

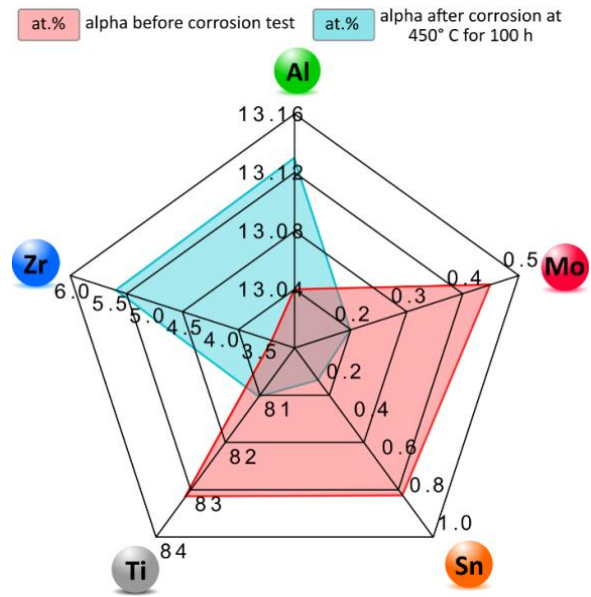

*Figure 3: Comparison of atom probe results for the $\alpha_p$ composition before corrosion cracking (red) and after corrosion cracking, near the crack tip (blue). Relatively large decreases in Mo and Sn are observed, and an increase in Zr content.*

Given the understanding that stress corrosion embrittlement in Ti alloys is caused by hydrogen, a very low concentration of D was unexpected. To complement the APT analyses, time-of-flight secondary-ion mass spectrometry (ToF-SIMS) spectra were also obtained from regions adjacent to the crack in samples maintained at cryo-temperature (see methods), Figure 4. The sample surface is covered by the corrosion deposits which hides the underlying microstructure. These analyses also showed substantial enhancement of $^{18}O$ in a ~20 µm region adjacent to the crack wall, far higher than the $^{18}O$ concentration measured in the rest of the sample. We have observed both $^{18}O$ and $^{16}O$ peaks, however, only the $^{18}O$ peak is shown for clarity to eliminate $^{16}O$ counts coming from the atmospheric oxygen. In contrast, whilst some D was measured, this was at a level over 1000 times lower than for $^{18}O$ and barely higher than for the $H_2^+$ ion usually found in the ultrahigh vacuum system.

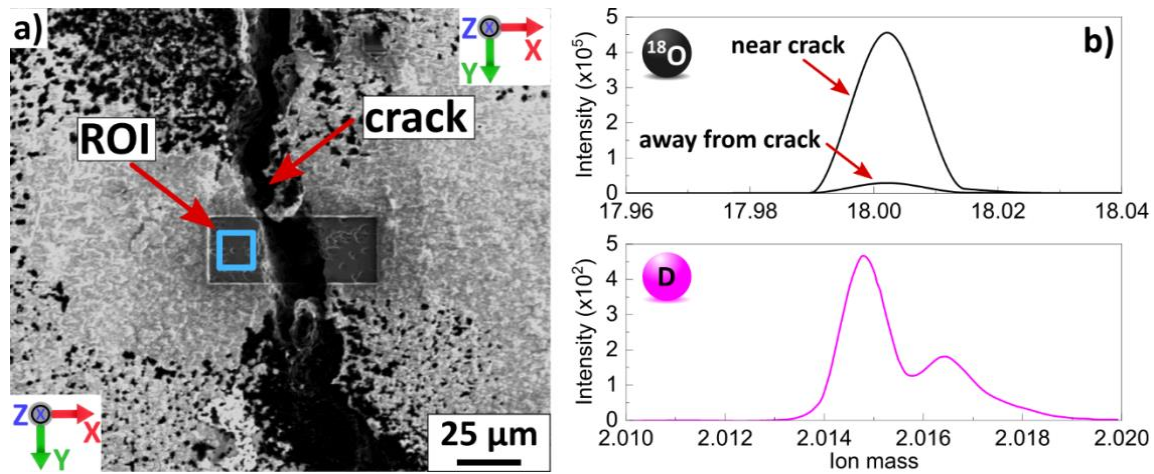

*Figure 4: ToF-SIMS analysis adjacent to crack. (a) Scanning electron microscopy image showing the region from which ToF-SIMS spectrum were captured. (b) Oxygen and deuterium mass spectra obtained near the crack.*

It is also possible to measure the presence of interstitial O and H using EELS in the TEM, Figure 5. The EELS analysis was done on the crack tip region, Figure 5a, where most of the original material was consumed during the corrosion reactions. This showed a prominent oxygen loss peak at 532 eV near the crack tip. It is difficult to confirm the presence of D/H from EELS since the electron energy loss peak of low atomic number elements overlaps with the plasmon peak, but at minimum EELS does not suggest that a large concentration of D/H was present. We do not claim here – yet – that we can completely exclude the possibility of hydrogen embrittlement in the present case, but the low concentration of H/D compared to O begs the question of whether it actually plays the main role in the embrittlement of the matrix (as distinct from the surface) during Ti stress corrosion cracking.

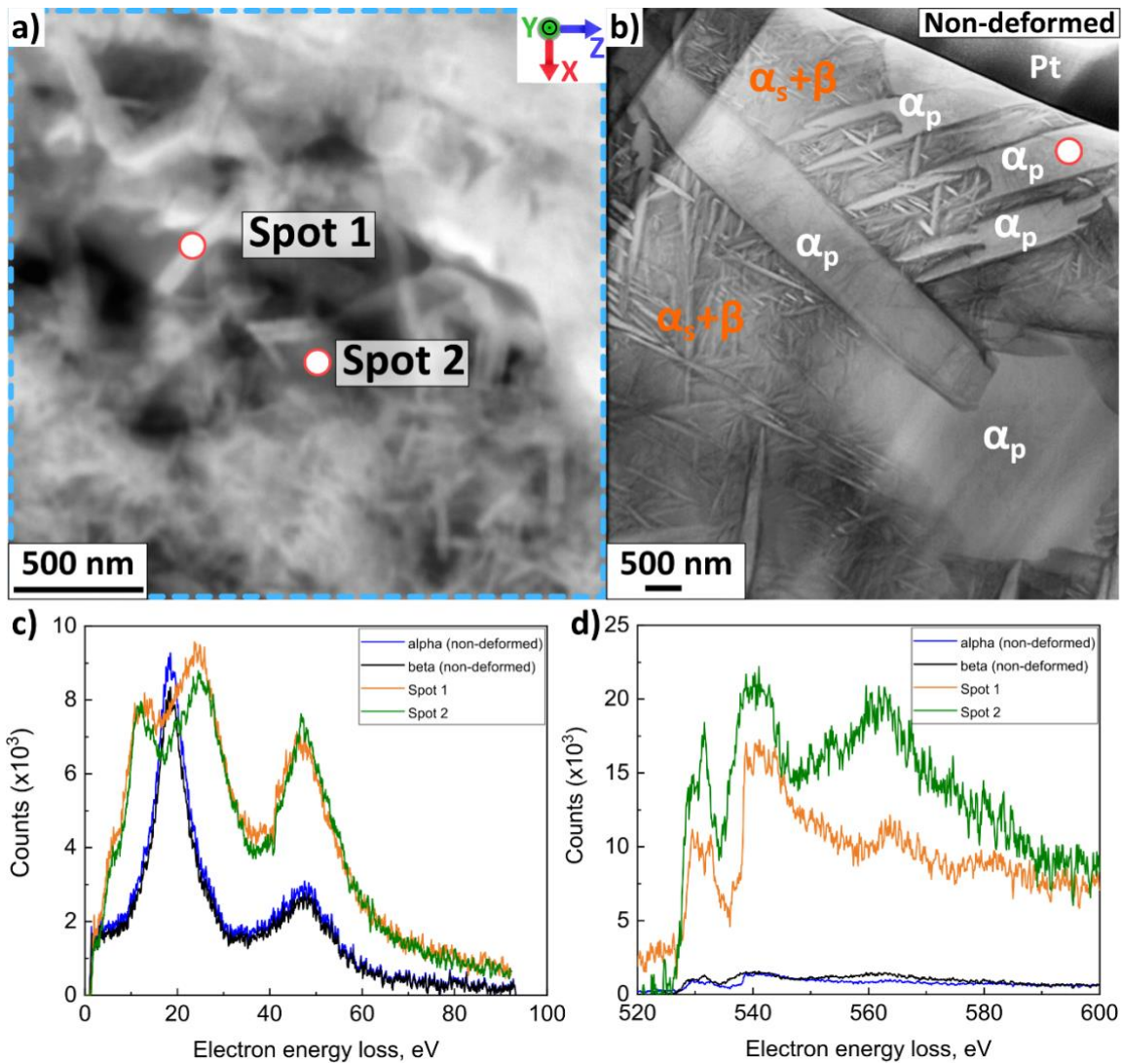

*Figure 3: EELS analysis for deformed and non-deformed Ti-6Al-2Sn-4Zr-6Mo alloy. (a) The regions near the crack tip from where the EELS spectrum were captured. (b) TEM lift-out from the as-received sample showing the region where EELS spectra were captured. (c) Comparison of low loss energy spectrum between tested and as-received sample. (d) Core loss spectrum showing the oxygen loss in the sample tested.*

### 3.4 Ti$_3$Al formation and embrittlement.

To better understand the cracking process and its interplay with local composition, a TEM specimen was prepared from the crack tip, Figure 6a (see methods for details). The analysis was done on the $\alpha_p$ grains of this TEM foil, Figure 6b. Ti$_3$Al precipitates, on the order of few nm in size, could readily be observed near the crack in dark field imaging with the superlattice reflection (shown with a circle in the inset of the figure), Figure 6c. High-resolution imaging was performed, Figure 6d, and FFT pseudo-diffraction patterns were obtained from those precipitate. An FFT from the matrix without precipitates is also shown for comparison. O reduces the solubility of Al in $\alpha$-Ti, leading to local clustering. Dark field imaging on the sample far away from the crack also showed Ti$_3$Al precipitation, albeit with an apparently lower number density.

The observation of precipitates near and far away from the crack suggests both O ingress and the heat treatment, combined with the alloy bulk Al and O content are responsible for this precipitation. Then ordering of these aluminium rich regions lowers the energy of the system and leads to the formation of the α₂ precipitates[36].

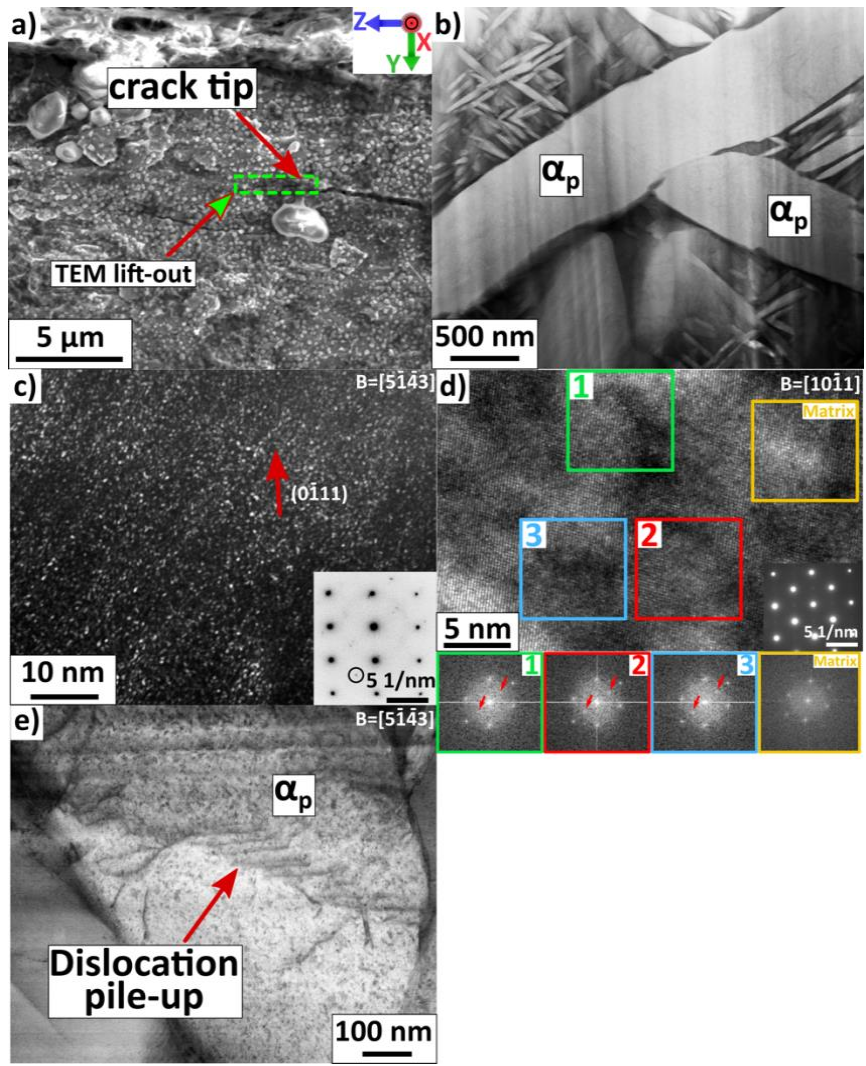

*Figure 4: TEM analysis at the crack tip. (a) SEM image showing the location from where the TEM lift-out was made ahead of the crack tip. (b) STEM BF image showing the region analysed in the foil. (c) TEM dark field image showing nanosized Ti₃Al precipitates in the primary alpha phase. (d) HRTEM image showing the precipitates and 1-4 the corresponding FFT with superlattice spots in 1-3. (e) BF-STEM image showing <a> basal pile-up under B = [5 $\bar{1}$ $\bar{4}$3].*

BF-STEM imaging revealed the presence of a slip band piled-up near the crack at an $\alpha_p$ grain, Figure 6(f), which was found to be composed of basal <a> dislocations by g.b invisibility analysis under two beam conditions. The $\alpha_2$ Ti₃Al phase is an ordered *hcp* superstructure of the $\alpha$ phase and therefore dislocations are required to traverse it in pairs, which reduces the propensity for cross slip and hence, forest hardening. Slip bands are therefore a flow localisation mechanism that can produce high stress along slip bands, leading to crack propagation. Previously, oxygen associated with oxidation of a titanium surface has been shown to cause Ti₃Al precipitation[37], consistent with the ternary Ti-Al-O phase diagram. The influence of the level of oxygen on slip planarity in Ti-Al alloys was also reported by Gray et al.[38]. Others also observed that dislocations, e.g. those associated with a crack tip plastic zone, can enhance oxygen ingress into titanium surfaces[39].

## 4 Discussion

Overall, the combination of APT, ToF-SIMS and EELS analyses on the one hand, and the observation of Ti₃Al precipitates, on the other hand, force us to reconsider the embrittlement mechanism. Our results suggest that a substantial ingress of oxygen, of >5 at.%, occurred in the α phase locally, around the crack tip, and that high hydrogen concentrations near the crack tip were not observed. Recent work of ours has shown that H can be detected, and implantation avoided, where a cryogenic atom probe preparation workflow is used[9,15]. The solubility of oxygen in α-Ti is the highest of all elements at around 33 at.%[40], much higher than in β due to the smaller size of β tetrahedral sites than octahedral α sites. Dislocations in the β were observed to be decorated with both Zr (17 at.%) and $^{18}$O (12 at.%). However, the two solutes do not always co-segregate at the same segment of the dislocation. Figure 2c shows this within a region of interest normal to the dislocation line. Figure 7a is a close-up on Figure 2a, and Figure 7b highlights the different distribution of O and Zr along the dislocation line on maps calculated from the voxelized APT data (1x1x1 nm, with a 3x3x1.5 Gaussian smoothing). The composition measured at dislocations in β contrast at the α/β interface that are decorated with Zr, but no elevated O levels were found. This supports the hypothesis that mobile dislocations in β may have provided the transport pathways for O through the alloy.

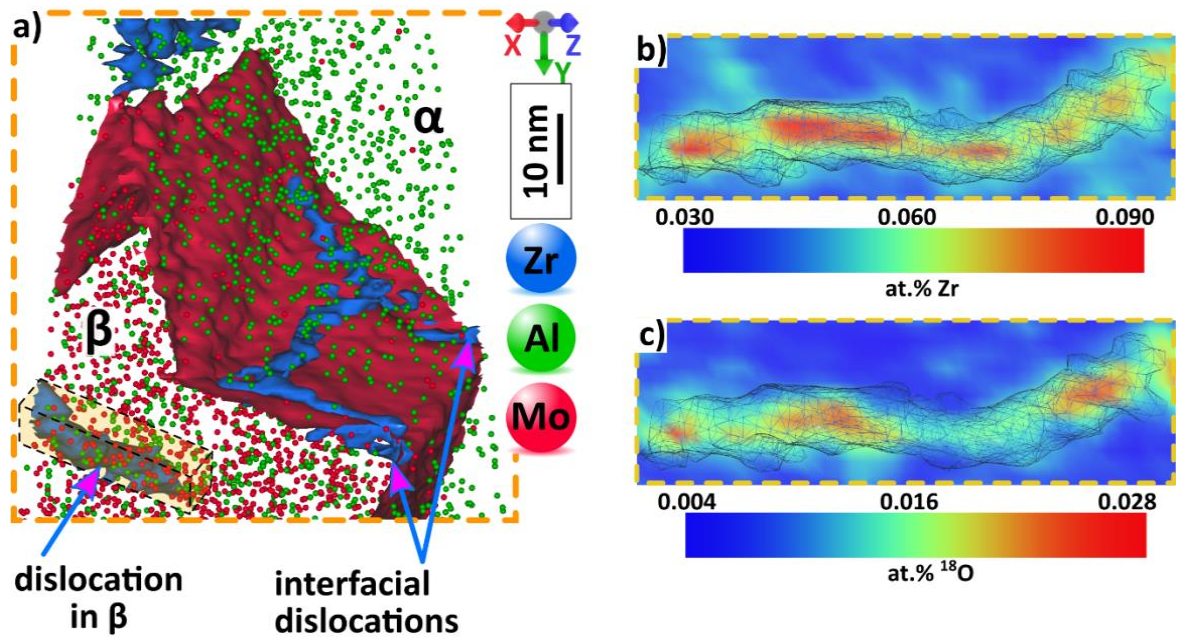

*Figure 7: Detail from the APT reconstruction in Figure 2α, showing the interfacial dislocations and dislocation in the beta phase. b) and c) 2D contour map of Zr and $^{18}$O for the dislocation in beta as denoted by the yellow sampling box at the bottom-left of (a).*

This leads to the question of why O from water vapour is split (e.g. from the hydroxide ion) and then deposited during the corrosion reactions. First, we note that a number of alloy chlorides $MCl_x$ are possible, and that when these are produced on reaction with HCl and $H^+$-$OH^-$ water complexes in the overall reaction sequence, it is perfectly possible that O might be implanted in solution; indeed, it is in contact with oxide corrosion product and therefore the alloy should be at its solubility limit at the interface. This is supported by the observation that the alloy composition depletes in high-volatility, reactive chloride-formers such as Mo and Sn in the vicinity of the crack tip. In addition, the formation of titanium dioxide is thermodynamically favourable, in comparison to that of titanium dihydride or even water, with enthalpies of formation of -940, -144 and 285 kJ/mol, respectively[30], and high-temperature oxidation of Ti involves the in-diffusion of O[41]. It is also notable that rutile $TiO_2$ is commonly used for water splitting[7]. The presence of O at the surface and in the subsurface region is therefore unavoidable, and its penetration inside the material during surface oxidation in the presence of water vapour has been previously reported[42]. The higher chemical potential of O at the surface compared to the bulk, as well as the possibility for transport by diffusion through α or accelerated via pipe diffusion through β makes ingress of O inside the alloy even more likely. Our results suggest that there in contrast, hydrogen originating from the dissociation of $D_2^{18}O$ either forms hydrides on the surface (which we have observed), penetrates and subsequently diffuses away into the beta (which is effectively an infinite sink for a large component and a small corrosion site), or simply recombines at the surface. A possible schematic is provided in Figure 8.

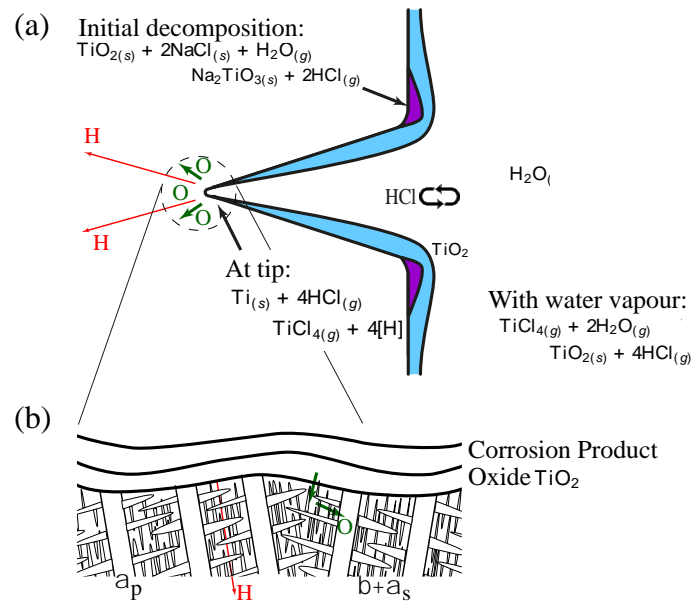

*Figure 8: Schematic of the corrosion implantation process. (a) showing the initial decomposition of the NaCl and then cycling of Cl- to crack water vapour in the gas stream and implant solute [O] and [H] (adapted from [8]), and (b) at the crack tip showing diffusion away through the β matrix of the implanted [H], with low solubility in the α; and via pipe diffusion of solute [O] in β to then enter and embrittle the α phase.*

There is also a question around kinetics of stress corrosion and hydride cracking mechanisms. In delayed hydride cracking type scenarios, e.g in zirconium alloys, the crack growth rate is limited by the rate of hydride formation and cracking ahead of the crack tip. In our experience of NaCl and AgCl SCC in components, typically a transition is observed to conventional low cycle fatigue at longer crack lengths (and higher ΔK), which reflects a stagnation in the supply of corrodant. This is consistent with the relatively slow kinetics of oxygen diffusion in titanium surfaces.

Such elevated O contents could embrittle the material by: i) increasing the friction stress for dislocation motion ii) promoting cross-slip on the pyramidal plane[43] and iii) segregating on the octahedral sites along the [$1\bar{2}13$] direction which are coplanar with the first order pyramidal plane [44]. In addition, the $Ti_3Al$ precipitates and D at the crack tip enhances extensive localized plasticity by dislocation pile-up and increasing dislocation velocity by HELP mechanism respectively causing crack propagation. The hydrogen shielding of the interaction of dislocations with elastic stress centres can account for the observed hydrogen-enhanced dislocation mobility and the shielding effect arises from the stress fields of point defects. The hydrogen reduces the repulsive interaction between the dislocations of same sign and the attractive force between the dislocations of opposite sign. Once the crack advances the hydrostatic stress state near the new crack front attracts both O and D atoms and further crack growth occurs by O embrittlement and extensive plasticity by D and $\alpha_2$ precipitates.

## 5 Conclusion

To conclude, our new set of experimental data shed new perspective on the stress corrosion embrittlement of titanium alloys by halides. We demonstrated a strong ingress of oxygen in critical parts of the microstructure, and only minute amounts of hydrogen/deuterium. Such a small ingress of hydrogen can still have dramatic effects on the local plasticity, however, the effect of oxygen on the microstructural evolution, and compositional variation, in part associated to plasticity-assisted diffusion, had so far not been reported as clearly in stress-corrosion cracking condition. The precipitation of an ordered phase triggered or facilitated by the high oxygen content, and the associated slip localisation, are poised to accelerate the crack propagation. We note that chloride and iodide stress corrosion is a general problem in metals, e.g. with I-induced stress corrosion in Zr in nuclear plant, and therefore suggest that O-implantation and associated embrittlement of crack tips at the nanoscale may also be relevant to halide stress corrosion cracking more generally.

**Acknowledgements**. At Imperial, the support of Dr C McGilvery and Dr RJ Chater are gratefully acknowledged. Work at Imperial was funded by EPSRC (EP/T01041/X, EP/L025213/1, EP/K034332/1) and Rolls-Royce plc. BG acknowledges financial support from the ERC-CoG-SHINE-771602. We are grateful for the financial support from the BMBF via the project UGSLIT and the Max-Planck Gesellschaft via the Laplace project. We thank Uwe Tezins & Andreas Sturm for their support to the FIB & APT facilities at MPIE.

**Author Contribution Statement.** DD & BG designed the study and drafted the article. SJ & YS performed the bending and corrosion tests. SJ performed the TEM and SIMS analyses. PK & YC performed the APT, PK & BG reconstructed and processed the APT data. All authors commented on the draft.